\documentclass[global,twocolumn]{svjour}
\usepackage[british]{babel}
\usepackage{graphics}
\usepackage{amsmath}
\usepackage{amssymb}
\usepackage{varioref}
\usepackage{hyperref}
\usepackage{bibentry}
\usepackage[square,numbers,sort&compress]{natbib}

\newcommand{\ket}[1]{\lvert#1\rangle}

\newcommand{\avs}[1]{\langle  #1  \rangle}

\usepackage{mathtools}

\makeatletter
\newlength{\@tmpwideincludespace}
\newcommand{\wideincludegraphics}[1]{%
  \setlength{\@tmpwideincludespace}{\textwidth+\marginparsep+\marginparwidth}
  \ifthispageodd{%
    \makebox[0pt][l]{%
      \includegraphics[keepaspectratio,%
        width=0.999\@tmpwideincludespace,totalheight=0.999\textheight]{#1}%
    }%
    \hfill%
  }{%
    \hfill%
    \makebox[0pt][r]{%
      \includegraphics[keepaspectratio,%
        width=0.999\@tmpwideincludespace,totalheight=0.999\textheight]{#1}%
    }%
  }%
  \clearpage%
}
\makeatother

\makeatletter
\newcommand*{\@titleimage}{}
\newcommand*{\titleimage}[1]{\gdef\@titleimage{#1}}
\makeatother

\journalname{Applied Physics B: Lasers and Optics}
\begin{document}
\sloppy
\title{Single-site- and single-atom-resolved measurement of correlation functions}
\author{M. Endres\inst{1} \and M. Cheneau\inst{1} \and T. Fukuhara\inst{1} \and C. Weitenberg\inst{1,2} \and P. Schau\ss\inst{1}\and C. Gross\inst{1} \and L. Mazza\inst{1,3} \and M.C.~Ba\~nuls\inst{1} \and L. Pollet\inst{4}, I. Bloch\inst{1,4}\and S. Kuhr\inst{1,5}
}                     
%
%
\institute{Max-Planck-Institut f\"ur Quantenoptik, 85748 Garching, Germany \and 
Laboratoire Kastler Brossel, CNRS, UPMC, Ecole Normale Sup\'{e}rieure, 75005 Paris, France \and Scuola Normale Superiore, 56126 Pisa, Italy \and Ludwig-Maximilians-Universit\"at, 80799 M\"unchen, Germany \and University of Strathclyde, Department of Physics, SUPA, Glasgow G4 0NG, United Kingdom}
%
%
\maketitle
\begin{abstract}
Correlation functions play an important role for the theoretical and experimental characterization of many-body systems. In solid-state systems, they are usually determined through scattering experiments whereas in cold-gases systems, time-of-flight and in-situ absorption imaging are the standard observation techniques. However, none of these methods allow  the in-situ detection of spatially resolved correlation functions at the single-particle level. Here we give a more detailed account of recent advances in the detection of correlation functions using in-situ fluorescence imaging of ultracold bosonic atoms in an optical lattice. This method yields single-site and single-atom-resolved images of the lattice gas in a single experimental run, thus gaining direct access to fluctuations in the many-body system. As a consequence, the detection of correlation functions between an arbitrary set of lattice sites is possible. This enables not only the detection of two-site correlation functions but also the evaluation of non-local correlations, which originate from an extended region of the system and are used for the characterization of quantum phases that do not possess (quasi-)long-range order in the traditional sense.\\
\end{abstract}
\section{Introduction}
The use of ultracold atoms for the study of strongly interacting many-body systems has undergone remarkable development in recent years \cite{Bloch:2012}. Prominent examples include the achievement of the strongly interacting regime of bosonic and fermionic gases in optical lattices  \cite{Greiner:2002a, Paredes:2004, Bloch:2008c, Jordens:2008, Schneider:2008} and studies of the BEC-BCS crossover by means of Feshbach resonances \cite{Randeria:2012}. The success of this approach is based on the high degree of control that has been achieved over the system parameters. In particular, the underlying Hamiltonian is usually known and its parameters, such as the interaction strength or the effective mass, can be accurately determined and tuned over a large range. Furthermore, ultracold gases experiments offer versatile detection techniques, such as in-situ absorption,  in-situ phase-contrast, and time-of-flight imaging \cite{Shin:2008,Gemelke:2009,Nascimbene:2010,Bloch:2008c}. The latter has been combined with spectroscopic techniques, such as momentum-resolved radio-frequency spectroscopy and momentum-resolved Bragg spectroscopy \cite{Stewart:2008, Ernst:2009}, to gain access to the excitation spectrum of the many-body system.\\
Recent advances in the high-resolution in-situ fluorescence imaging of atoms in optical lattices \cite{Bakr:2009, Bakr:2010, Sherson:2010} have pushed detection capabilities to the fundamental level of individual atoms. Specifically, this technique allows for the single-site-resolved detection of a lattice gas in the Bose-Hubbard regime \cite{Fisher:1989,Jaksch:1998}.
The Bose-Hubbard Hamiltonian is given by
\begin{align}
\hat H_{\text{\rm{BH}}}=-J \sum_{\langle i, j \rangle} \hat{a}^\dagger_j \hat{a}_i+\frac{U}{2}\sum_{i} \hat{n}_i (\hat{n}_i-1)- \sum_i \mu_i \hat{n}_i,
\label{equ:bose_hubbard_simple}
\end{align}
where $\hat{a}^\dagger_i$ ($\hat{a}_i$) is the boson creation (annihilation) operator on lattice site $i$, $\hat{n}_i=\hat{a}^\dagger_i \hat{a}_i$ is the boson number operator, $J$ is the hopping matrix element, $U$ is the on-site interaction energy,  $\mu_i$ is the local chemical potential,  and the first sum runs over all nearest neighbors. The competition of the hopping and interaction process gives rise to a quantum phase transition between a superfluid and a Mott-insulating phase \cite{Fisher:1989, Jaksch:1998, Sachdev:2011}.\\
The high-resolution fluorescence experiments \cite{Bakr:2009, Bakr:2010, Sherson:2010} are able to detect the on-site parity, ${\rm mod}_2\, n_i$, where $n_i$ is the occupation number of site $i$ (see Sec.\,\ref{setup}). The average on-site parity has been used to study the superfluid-Mott insulator transition \cite{Bakr:2010} and to develop an in-situ temperature measurement in the deep Mott-insulating limit\,\cite{Sherson:2010}. The latter has recently been combined with amplitude-modulation spectroscopy, forming a sensitive tool for the study of the many-body excitation spectrum, which was used to detect a `Higgs' amplitude mode close to the superfluid-Mott-insulator transition \cite{Endres:2012, Pollet:2012}. Furthermore, an antiferromagnetic quantum Ising spin chain has been simulated in a tilted optical lattice, where nearest-neighbor spin correlations map onto the average on-site parity \cite{Simon:2011}, and an orbital excitation blockade could be directly observed \cite{Bakr:2011}.\\
The high-resolution technique can also be used to control individual atoms and to change Hamiltonian parameters on the level of individual lattice sites. In particular, it has been experimentally demonstrated that the spin of individual atoms in a Mott insulator can be addressed \cite{Weitenberg:2011}. This technique has recently been employed to study a mobile spin impurity in a strongly interacting one-dimensional (1d) system \cite{Fukuhara:2013}.\\
In addition to the average on-site parity, high-resolution in-situ fluorescence imaging can be used to evaluate correlations between different lattice sites. We used this possibility to detect two-site and non-local correlations across the superfluid-Mott-insulator transition \cite{Endres:2011} and to image the spreading of correlations after a sudden change of the Hamiltonian parameters in the Mott-insulating regime \cite{Cheneau:2012}.\\
Here we give a more detailed account of several technical and physical aspects of the detection of two-site and non-local correlations with a focus on the equilibrium situation in 1d \cite{Endres:2011}. In particular, we give a self-contained description of the core experimental apparatus and sequence (Sec.\,\ref{setup}). Following this, we discuss the observable and the limitation of the detection technique in more detail (Sec.\,\ref{paritysection} and \ref{observable}). Furthermore, we give a short introduction to the single-site-resolved detection of Mott insulators in the atomic limit \cite{Sherson:2010} (Sec.\,\ref{atomic_limit}) as far as it is needed for the understanding of the following chapters. After a general introduction to correlation functions (Sec.\,\ref{correlation_functions}), we demonstrate how two-site correlation functions can be used for the detection of correlated particle-hole pairs (Sec.\,\ref{two_point}). We give a more detailed introduction to two-site parity correlation functions and particle-hole pairs than in Ref.\,\cite{Endres:2011} and show additional experimental data for two-site correlations as a function of the in-trap position. We continue with a detailed analysis of non-local order in the Bose-Hubbard model (Sec. \ref{non_local}) for which we show a pertubative result that we compare to numerical calculations. We finish with a short description of the experimental results for non-local correlations where our focus is on the role of three-site correlations.\\
\section{Introduction to single-site- and single-atom-resolved detection}
\subsection{Experimental setup and sequence}\label{setup}
\subsubsection{State preparation}
The starting point of our experiments was a two-dimensional (2d) degenerate quantum gas consisting of several hundred $^{87}$Rb atoms in the hyperfine state $5S_{1/2}$, $F=1$, $m_F=-1$, which was prepared in a single anti-node of an optical lattice in the vertical direction (see Fig.\,\ref{fig:1}a). For details concerning the preparation procedure, we refer to the supplementary material of Ref.\,\cite{Endres:2011}.\\
The vertical lattice was generated by  interference between an incoming laser beam and its reflection from a vacuum window, which has a high-reflectivity coating for the laser wavelength $\lambda_{\rm L}=1064\,\rm{nm}$, and the lattice spacing in the vertical direction was $a_{\rm lat}=\lambda_{\rm L}/2=532\,\rm{nm}$. The laser frequency was red-detuned to the D2 and D1 line, yielding an attractive ac Stark potential \cite{Grimm:2006}. The vertical lattice depth $V_z$ was kept constant at approximately $21\,E_r$, where $E_r$ denotes the lattice recoil energy $E_{r}=h^2/(8m a_{\rm lat}^2)$ with $m$ being the atomic mass of $^{87}$Rb. Due to this tight confinement and an additional energy offset from gravity, tunneling of the atoms in the vertical direction was negligible for the duration of the experiment. Additionally, the vertical confinement was much stronger than the harmonic confinement in horizontal direction that results from the Gaussian beam shape. In this pancake-like geometry, the atom cloud formed a 2d degenerate Bose gas (see, e.g., Ref.\cite{Hadzibabic:2008}).\\
Subsequently, the gas was loaded into a 2d optical lattice formed by two optical lattice axes in the horizontal direction, which were created by reflections from mirrors outside of the vacuum chamber. The lattice spacing in both directions was $a_{\rm lat}=\lambda_{\rm L}/2=532\,\rm{nm}$ and the axes intersected at $90^\circ$, resulting in a simple 2d square lattice. For the loading, the lattice depths were increased to values between $5\,E_r$ and $23\,E_r$, following s-shaped functions with durations of $120\,\rm{ms}$ (Fig.\,\ref{fig:1}b). These lattice ramps were quasi-adiabatic, i.e. slow enough to keep the system close to its many-body ground state, as can be seen by the low defect density in the system (Sec.\,\ref{atomic_limit}).
\subsubsection{Fluorescence imaging} 
\begin{figure*}
\resizebox{1\textwidth}{!}{%
\includegraphics{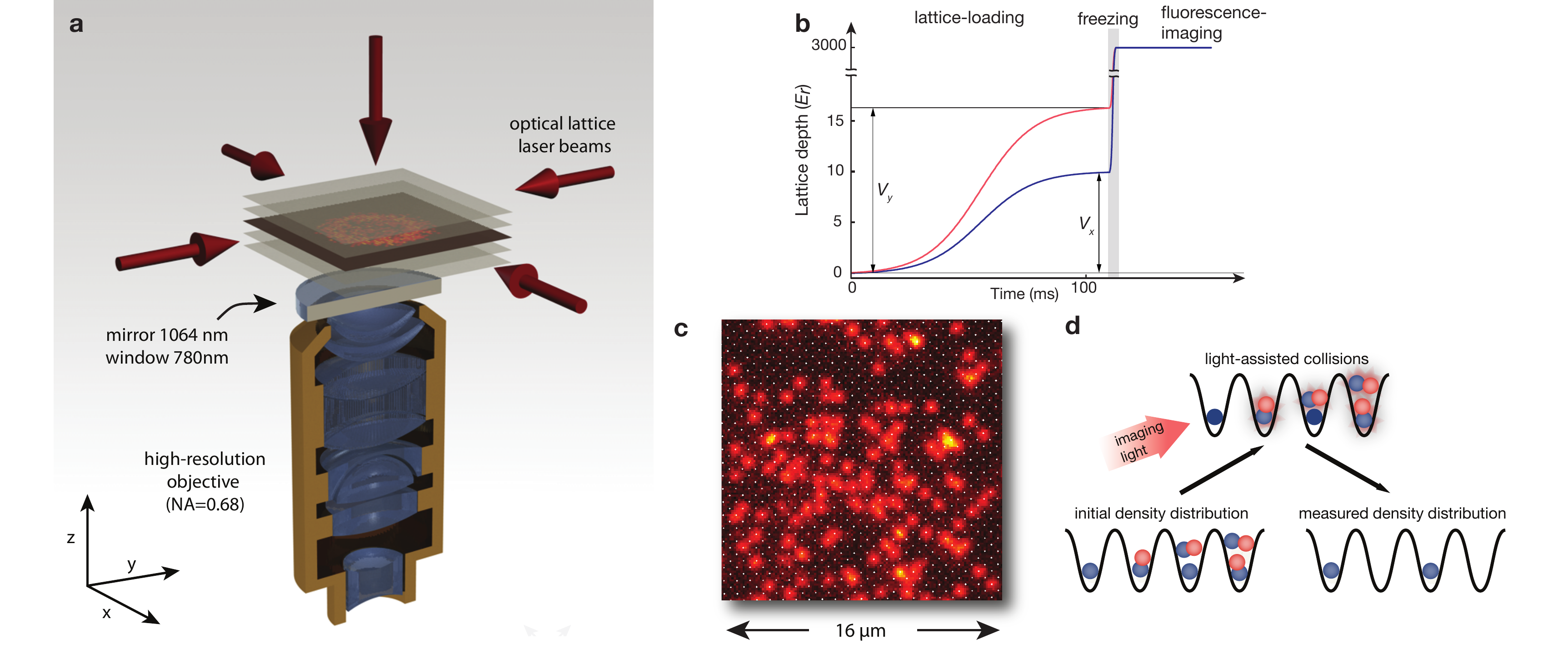}
}
\caption{\textbf{Experimental setup and fluorescence imaging. a}, Schematic image showing the optical lattice geometry, the 2d quantum gas and the high-resolution objective. The optical lattice setup consisted of two horizontal lattice axes (in the $x$ and $y$-directions) and a vertical lattice axis (in the $z$-direction). The latter was created by reflection from a vacuum window with a coating that reflects the lattice wavelength ($1064 \,\rm{nm}$) and transmits the imaging wavelength ($780\,\rm{nm}$). The 2d quantum gas was prepared in a single anti-node of the vertical lattice and is imaged with the high-resolution objective using fluorescence detection. \textbf{b}, Horizontal lattice depth during a typical experimental sequence. The loading into the square lattice is achieved by quasi-adiabatic s-shaped ramps with a duration of $120\,\rm{ms}$. The final lattice depths in $x$- and $y$-directions $V_x,V_y$ were independently controlled. For the detection, the lattice gas is frozen by rapidly increasing the lattice depths to approximately $3000\,E_r$. \textbf{c}, Typical fluorescence image of a dilute thermal cloud as seen on an EMCCD camera. Single atoms are visible with a high signal-to-noise ratio. White dots mark the sites of the 2d lattice created by the horizontal lattice axes.  \textbf{d}, Illustration of the parity-projection mechanism. Figs. \textbf{a} and \textbf{c} are from Ref.\,\cite{Sherson:2010}.}
\label{fig:1}       
\end{figure*}

We detected the atoms in-situ using fluorescence imaging\,\cite{Sherson:2010,Weitenberg:2012}. The imaging was performed with a high-resolution objective, which was placed in front of the vacuum window that is also used as a mirror for the vertical lattice axis. The numerical aperture of the objective was ${\rm NA}=0.68$, yielding a spatial resolution of about $700\,\rm{nm}$ for an imaging wavelength of $780\,\rm{nm}$. \\
For the imaging, all lattice depths were increased to typically $3000\, E_r\approx k_B \cdot 300\,  {\rm \mu K}$  on a time scale much faster than the many-body dynamics (Fig.\,\ref{fig:1}b).  As a result, the original density distribution of the gas was instantaneously frozen. The detailed sequence for the freezing consisted of a two-step process, where the first step was an exponential ramp to  $\approx 80 \,E_r$ with a duration of $0.2\,$ms that already fully suppressed the dynamics (not shown in Fig.\,\ref{fig:1}b). The second ramp to $\approx 3000\, E_r$ is included to have sufficiently deep lattices for the imaging process.\\
To image the frozen gas, we optically pumped the atoms from $5S_{1/2}$, $F=1$ to $5S_{1/2}$, $F=2$ and shone in imaging laser light with a red-detuning of $\approx 50\,$MHz to the free-space resonance of the $5S_{1/2}$, $F=2$ to $5P_{3/2}$, $F=3$ transition. The scattered photons were observed through the high-resolution objective and an additional lens, which focused the light on an EMCCD (electron-multiplying charge-coupled device) camera.\\
Typical fluorescence images show individual atoms in the 2d lattice (see Fig.\,\ref{fig:1}c). We first discuss a dilute, non-degenerate cloud and turn to dense, degenerate clouds in Sec.\,\ref{atomic_limit}. Approximately $5000$ photons per atom are detected during an illumination time of about $1\,{\rm s}$, resulting in a high signal-to-noise ratio.  To achieve this, we had to suppress tunneling (or even loss) of the atoms during the illumination time. Such tunneling processes can occur even in a very deep optical lattice of $3000\, E_r$ if the atoms are thermally excited to higher bands as a result of the recoil heating by the imaging light. To suppress this thermally activated hopping, we laser-cooled the atoms by setting up the imaging beams in an optical molasses configuration, which resulted in sub-Doppler temperatures of $\approx 30\,\mu {\rm K}$.\\
Note that the imaging process is destructive for the many-body state (see Sec.\,\ref{observable} for details). To collect statistics for a given observable, the experiment had to carried out repeatedly, starting with the preparation of the 2d degenerate gas.
\subsection{Parity projection and parity operator}\label{paritysection}
Our images show the parity of the occupation number on each lattice site (Fig.\,\ref{fig:1}d). This is a consequence of a pairwise loss process due to light-assisted collisions \cite{Weiner:1999, Bakr:2010, Sherson:2010} occurring on a time scale of typically $100\, \mu s$, which is much faster than the illumination time of about $1\,s$. Without additional steps, this loss would occur in the beginning of the illumination time, and some of the lost atoms would be recaptured due to the molasses cooling, leading to an unwanted background signal.  To avoid this, we trigger light-assisted collisions before repumping and before switching on the imaging light with a 50\,ms pulse of light resonant for the $5S_{1/2}$, $F=2$ to $5P_{3/2}$, $F=3$ transition. At this stage in the sequence, the atoms are in the state $5S_{1/2}$, $F=1$ and the light is 6.8\,GHz red detuned for the $5S_{1/2}$, $F=1$ to $5P_{3/2}$, $F=2$ transition, but efficiently excites atom pairs into molecular states leading to a rapid loss. As a result, doubly occupied lattice sites appear as empty sites in the images. Similarly, three atoms on a site are imaged as a single atom as two atoms undergo a pairwise collision. In general,  the measured occupation number can be written as ${\rm mod}_2 \: n_i$ in terms of the actual occupation number $n_i$. For later use, we define an on-site parity operator $\hat{s}_i$ with eigenvalues $s_i$ that are $\pm 1$ for odd (even) parity:
\begin{align}\label{parity_operator}
\hat{s}_i\ket{n_i} =s_i\ket{n_i}= \left\{ 
  \begin{array}{l l}
    +1 \ket{n_i} & \quad \text{if $n_i$ is odd}\\
    -1 \ket{n_i} & \quad \text{if $n_i$ is even,}
  \end{array} \right.
\end{align}
where $\ket{n_i}$ is an on-site Fock state with occupation number $n_i$ at site $i$.
\subsection{Description of the observable} \label{observable} 
\subsubsection{Interpretation as a projective measurement}
The described measurement technique detects more than just the average on-site parity but also captures the fluctuations and correlations in the system. An arbitrary many-body state $|\Psi\rangle$ (not necessarily in the atomic limit) at zero temperature can be written as a superposition of products of on-site Fock states as
\begin{align}
|\Psi\rangle=\sum_{\{n_i\}}\alpha_{n_1,...,n_N}|n_1,...,n_N \rangle,
\end{align}
where $|n_1,...,n_N \rangle=\prod_i |n_i\rangle$ and the sum runs over all possible configurations of on-site occupation numbers $\{n_i\}=(n_1,...,n_N)$.\\
The freezing of the density distribution and the subsequent scattering of imaging light can be interpreted as a projective measurement. It leads to a projection 
onto a specific state $|\Psi\rangle_{\rm proj}=|n_1,...,n_N \rangle$  with a quantum mechanical probability  $p(n_1,...,n_N)=|\alpha_{n_1,...,n_N}|^2$. 
\subsubsection{Observable without parity projection}
We first discuss the situation as it would be without parity projection. The crucial point is that the measurement would yield information about all occupation numbers $n_i$ of the projected state $|\Psi\rangle_{\rm proj}$ in a single experimental run. In each iteration of the experiment, we would observe a new set of occupation numbers corresponding to a different $|\Psi\rangle_{\rm proj}$. In this way, we could gather an increasing amount of statistics and finally reconstruct the full joint probability distribution $p(n_1,...,n_N)$ to observe a specific set of occupation numbers $(n_1,...,n_N)$ on all lattice sites. This includes more information than the average on-site density because from the knowledge of $p(n_1,...,n_N)$, one could calculate all possible density-density correlation functions $\langle \prod_{i \epsilon M}\hat n_i \rangle=\sum_{\{n_i\}}p(n_1,...,n_N)\prod_{i \epsilon M }n_i$, where the product runs over an arbitrarily chosen set of lattice sites $M$.
\subsubsection{Observable including parity projection}
Including the parity-projection mechanism, the most general observable is the joint probability distribution $p_p(s_1,...,s_N)$ to observe a set of on-site parities $(s_1  ,... , s_N)$ on all lattice sites, where the $s_i$ are eigenvalues of the on-site parity operator as defined in Eq.\,\ref{parity_operator}. However, the full reconstruction of $p_p(s_1,...,s_N)$ is a demanding task since it requires a very large number of experimental repetitions. In practice, we evaluated various correlation functions using an additional spatial average in order to reduce the statistical noise on our data.
\subsubsection{Limitation}
The detection technique can not directly distinguish the pure state $|\Psi\rangle=\sum_{\{n_i\}}\alpha_{n_1,...,n_N}|n_1,...,n_N \rangle,$ from a mixed state described by the density operator
\begin{align}
\hat \rho=\sum_{\{n_i\}}|\alpha_{n_1,...,n_N}|^2 \hat{P}_{ |n_1,...,n_N \rangle},
\end{align}
with the projection operator $\hat{P}_{ |n_1,...,n_N \rangle}= |n_1,...,n_N \rangle \langle n_1,...,n_N |$. The reason is that both of these states yield the same joint probability distributions $p(n_1,...,n_N)=|\alpha_{n_1,...,n_N}|^2$ and $p_p(s_1,...,s_N)$. Therefore, the imaging technique, with or without parity projection, can not detect off-diagonal elements (i.e., coherences) of density operators written in the on-site number basis.
\section{Single-site-resolved detection of dense, atomic-limit Mott insulators}
\label{atomic_limit}
\begin{figure}
\resizebox{0.5\textwidth}{!}{%
\includegraphics{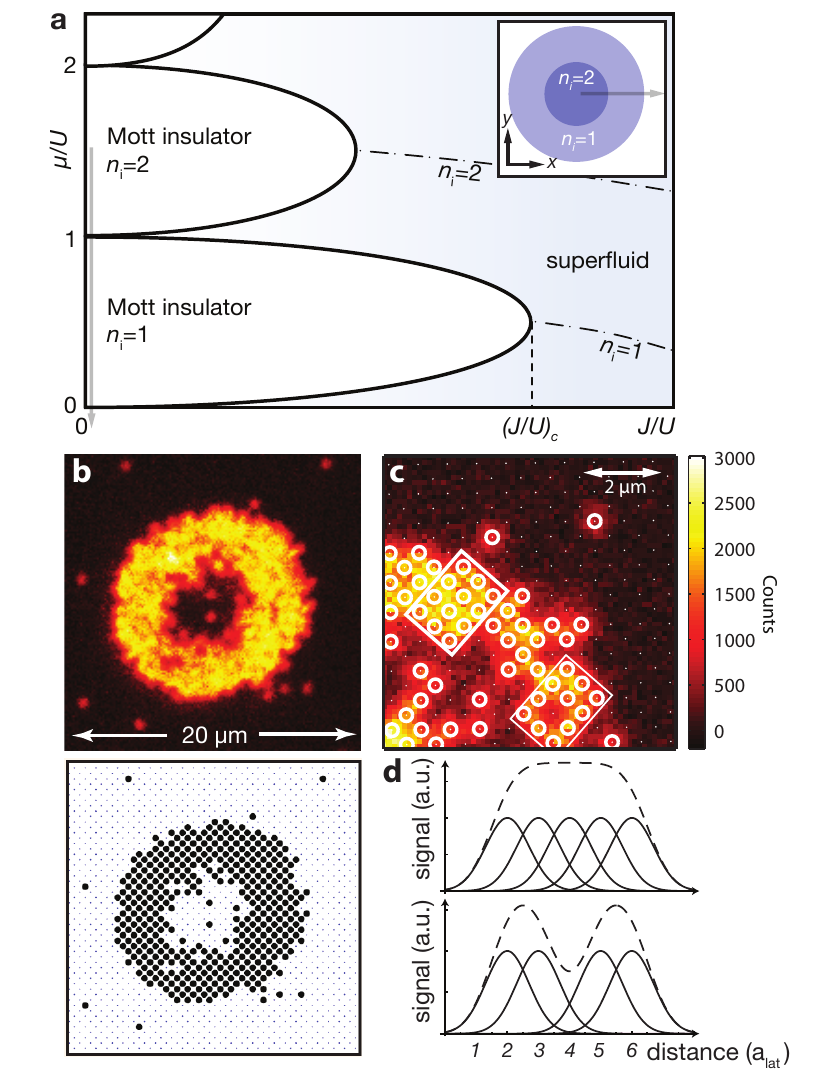}
}
\caption{\textbf{Imaging Mott insulators in the atomic limit. a,} Sketch of the phase diagram of the Bose-Hubbard model (Eq.\,\ref{equ:bose_hubbard_simple}) \cite{Fisher:1989}. For small $J/U$, the system tends to be in a Mott-insulating phase divided into lobes with constant average on-site occupation numbers $ n_i=\left\langle\hat{n}_i\right\rangle$ (white areas). For $J/U$ larger than a critical value $(J/U)_c$, the system is always superfluid (shaded blue area) and only lines with constant $n_i$ exist (dashed lines). Inset: Sketch of the local average density of a trapped system in the atomic limit ($J/U\approx0$) that can be considered as a cut in the Bose-Hubbard phase diagram along the $\mu$-direction (grey-shaded arrow). \textbf{b,}  Upper panel: experimental fluorescence image of a Mott insulator in the atomic limit with an occupation number of two in its center. Lower panel: Reconstruction of the atom distribution in the lattice. The sites in the central region appear empty due to the parity projection mechanism. \textbf{c}, Reconstruction in a typical high-density region. Open circles mark the positions of single atoms as found by the reconstruction algorithm. \textbf{d}, The reconstruction is possible even if the optical resolution (about $700\,\rm{nm}$) is slightly larger than the lattice spacing ($a_{\rm lat}= 532\,\rm{nm}$). Five atoms on neighboring lattice sites are sketched with their point spread functions (solid line, top panel). The sum of the signals (dashed line) shows only a broad feature. However, if one of the atoms is missing (lower panel) a dip in this feature appears. The reconstruction algorithm can therefore still distinguish the different occupations in the two boxed areas in \textbf{c}. Figs.\textbf{b,c} are from Ref.\,\cite{Sherson:2010}.}
\label{fig:2}       
\end{figure}
\subsection{Theoretical description of the atomic limit}
The atomic limit of the Mott-insulating phase ($J/U \approx 0$) yields a particularly simple description because, neglecting the tunneling term, the Bose-Hubbard Hamiltonian can be written as a sum of on-site terms as
\begin{align}
\hat H_{\text{\rm{BH}}}\approx\sum_i \hat H_i\equiv\sum_{i}\left[\frac{U}{2} \hat{n}_i (\hat{n}_i-1)-\mu_i\hat{n}_i\right].
\label{equ:atomic_limit}
\end{align}
The eigenstates of the local Hamiltonian $\hat H_i$ are on-site Fock states $|n_i\rangle$ with  eigenvalues ${E(n_i)=\frac{U}{2} {n}_i ({n}_i-1)-\mu_i {n}_i}$. The eigenstates of $\hat H_{\text{\rm{BH}}}$ are product states $|\Psi\rangle_{J/U=0}=\prod_i |n_i\rangle$ with eigenenergies $E=\sum_i E(n_i)$. The ground state of the system is therefore found by minimizing $E(n_i)$ on each lattice site. This yields $n_{0,i}=\lceil \mu_i/U \rceil$ for the local occupation number in the ground state, where $\lceil x \rceil$ is the ceiling of $x$. The ground state is then
\begin{align}
|\Psi\rangle_{0,J/U=0}=\prod_i |n_{0,i}\rangle.
\label{atomic_limit_ground}
\end{align}
One of the crucial features of this state is that it shows no number fluctuations, i.e., $\left<\hat{n}_i^2\right>-\left<\hat{n}_i\right>^2=0$. Number fluctuations only arise through thermal occupation of excited states by increasing or decreasing $n_{0,i}$. \\
In our system, the Gaussian shape of the lattice beams leads to a harmonic confinement in the center of the beams, and the local chemical potential is given by
\begin{align}
\mu_i=\mu-\frac{1}{2}m(\omega_x^2 x_i^2+\omega_y^2 y_i^2),
\end{align}
where $\mu$ is the global chemical potential and $\omega_x$, $\omega_y$ are the trapping frequencies in $x$,$y$-directions and $x_i$, $y_i$ denote the $x$,$y$-positions of lattice site $i$. We do not include a confinement term in the $z$-direction because our system is effectively 2d. The local chemical potential $\mu_i$, going outwards from the trap center, therefore follows a cut at $J/U\approx 0$ along the $\mu$ axis of the Bose-Hubbard phase diagram, starting at the global chemical potential $\mu$ and going to lower local chemical potentials (Fig.\,\ref{fig:2}a). As a consequence, the Mott lobes appear as concentric rings in a single experimental image.
\subsection{Single-site resolved detection in the atomic limit}
Experimentally, we reached the atomic limit by increasing the lattice depth of both horizontal lattices quasi-adiabatically (Sec.\,\ref{setup}) until tunneling between neighboring sites is completely suppressed. For the data presented in Fig.\,\ref{fig:2}, we used a final lattice depth of $V_x=V_y=23(1)E_r$  yielding $J/U\approx 3 \cdot10^{-3}$, which is about a factor of $20$ smaller than the critical value $(J/U)_c\approx 0.06$ for the superfluid-Mott-insulator transition in 2d \cite{Capogrosso:2008}.\\ 
After this preparation, we took in-situ fluorescence images of the system (Fig.\,\ref{fig:2}b). The pictures show regions with almost constant occupation number and low defect probability, consistent with the description in terms of on-site Fock states. Remaining number fluctuations are attributed to thermal excitations of the system (see Ref.\,\cite{Sherson:2010} for a detailed discussion).\\
In contrast to the image of a dilute thermal cloud shown in Fig.\,\ref{fig:1}c, single atoms are not easily distinguishable in the dense regions of the Mott-insulating shells. However, the signal-to-noise ratio and the spatial resolution of our imaging procedure are sufficient to reconstruct the atom distribution in the lattice using a computer algorithm (Figs.\,\ref{fig:2}c and d). The output of the algorithm is digital information in matrix-form containing the parity of the on-site occupation number for each lattice-site \cite{Sherson:2010, Weitenberg:2011}.

\section{Detection of correlation functions}
\subsection{Introduction to local and non-local correlation functions}\label{correlation_functions}
\subsubsection{Two-site correlation functions and local order parameters}
The following introduction to local and non-local order parameters follows in parts the discussion in Ref.\,\cite{Anfuso:2007a}. Local order parameters are based on two-site correlation functions
\begin{align}
C(d)=\langle \hat{A}_k\hat{B}_{k+d} \rangle,
\label{equ:local_order}
\end{align}
where $\hat{A}_k$ and $\hat{B}_{k+d}$ are, so far, not specified local operators acting at positions $k$ and $k+d$.  A system possesses long-range order in a two-site correlation function if 
\begin{align}
\lim_{d\rightarrow\infty}C(d)  =|C|^2 \neq 0.
\end{align}
If this is the case, $C$ is called a local order parameter. Typically, one studies the transition from an ordered phase, which possesses long-range order, to a disordered phase with \mbox{$C=0$}. Examples include off-diagonal long-range order in a Bose-Einstein condensate defined as $\lim_{d\rightarrow\infty}\langle \hat{\Psi}^\dagger_k \hat{\Psi}_{k+d}\rangle=|\Psi|^2\neq 0$, where $\Psi$ is the macroscopic wave function and $\hat{\Psi}^\dagger_k$ and $\hat{\Psi}_{k+d}$ are creation and annihilation operators. A simple example of magnetic order is ferromagnetic long-range order in the 1d transverse field Ising model, defined as
\begin{align} \label{simple_ferro}
\lim_{d\rightarrow\infty}\langle \hat{S}^z_k \hat{S}^z_{k+d} \rangle\neq 0,
\end{align}
where $\hat{S}^z_k$ is the $z$-component of the spin operator.\\
In a quasi-ordered phase, e.g., in low-dimensional systems, we often find an algebraic decay of $C(d)$ to zero and no finite value of $C$ (quasi-long-range order). A disordered phase is then characterized by a faster, exponential decay of $C(d)$.
\subsubsection{Multi-site correlation functions and non-local order parameters} 
Many phases that do not show \mbox{(quasi-)long-range} order in the above sense still show non-vanishing long-range correlations in multi-site correlation functions of the type
\begin{align}
O(l)=\langle \hat{A}_k \left( \prod_{k<m<k+l} \hat{u}_m\right) \,\hat{B}_{k+l} \rangle .
\label{eq:non_local_order}
\end{align}
This can be thought of as an extension of the two-site correlation function with a string of operators $\prod_{k<m<k+l}\hat{u}_m$ inserted between the endpoints. We call $O(l)$ a non-local correlation function because it gathers information about an extended region with length $l$. A system possesses non-local order if
\begin{align}
\lim_{l\rightarrow\infty}O(l)=O^2\neq 0.
\label{eq:non_local_order2}
\end{align}
This definition is a generalization of string order
\begin{align}
\lim_{l\rightarrow\infty}\langle  \hat{S}^z_{k} \left( \prod_{k< m< k+l}e^{i\pi \hat{S}^z_m } \right) \, \hat{S}^z_{k+l} \rangle \neq 0,
\label{string_spin1}
\end{align}
which was introduced in the context of spin-1 chains\,\cite{denNijs:1989,Kruis:2004}. This non-local correlation function detects a special type of anti-ferromagnetic order, which is hidden for a two-site correlation function (Fig.\,\ref{fig:4}a). Following Refs. \cite{Anfuso:2007a, PerezGarcia:2008}, we will use the term string order also for the more general definition in Eq.\, \ref{eq:non_local_order} and Eq.\,\ref{eq:non_local_order2}.\\
Another example is the disordered, paramagnetic phase in the 1d transverse field Ising model that shows a non-local order\,\cite{Kogut:1979}
\begin{align}
\lim_{l\rightarrow\infty}\langle\prod_{k\leq m \leq  k+d} \hat S_m^{x}\rangle \neq 0 ,
\end{align}
where $\hat S_k^{x}$ is the $x$-component of the spin operator at site $k$. This is in contrast to the ordered, ferromagnetic phase of the same model, which shows long-range order in a two-site correlation function based on $\hat S_k^{z}$ as defined in Eq.\,\ref{simple_ferro}.\\
We will argue in Sec.\,\ref{non_local} that 1d Mott insulators with unity filling possess a non-local order given by
\begin{align}
\lim_{l\rightarrow\infty} \langle \prod_{k\leq m\leq k+l}\hat s_m\rangle \neq 0
\label{eq:string1}
\end{align}
with the parity operator $\hat s_m$ defined in Eq.\,\ref{parity_operator}.
\\
This order parameter was introduced to study the transition from the Mott insulating phase to a Haldane insulating phase that can occur if long-range interactions are included in the Bose-Hubbard Hamiltonian\,\cite{DallaTorre:2006,Berg:2008}. The latter phase is characterized by an order parameter similar to the one in Eq.\,\ref{string_spin1}. In contrast, our focus is on the behaviour of the order parameter in Eq.\,\ref{eq:string1} at the transition from the Mott insulating to the superfluid phase within the standard Bose-Hubbard model (Eq.\,\ref{equ:bose_hubbard_simple}).\\
Further examples of systems that possess non-local order are spin-1/2 ladders \cite{Kim:2000} and fermionic Mott and band insulators \cite{Anfuso:2007b}. Recent theoretical studies explored the connection between non-local order and symmetries \cite{PerezGarcia:2008} and the connection to localizable entanglement \cite{Verstraete:2004,Verstraete:2004a,Popp:2005,Venuti:2005}. A measurement of the string order parameter of Eq.\,\ref{string_spin1} using spin-dependent lattice potentials has been proposed in Ref.\,\cite{Garcia_Ripol:2004}.
\subsection{Detection of two-site correlation functions and particle-hole pairs}
\label{two_point}
\subsubsection{Particle-hole pairs} In contrast to the atomic limit, Mott insulators at finite $J/U>0$ show number fluctuations even at zero temperature and are not described by simple product states as in Eq.\,\ref{atomic_limit_ground}. For $J/U \ll (J/U)_c$, these fluctuations appear in the form of correlated particle-hole pairs, formed by an extra particle and a missing particle on two nearby sites. In the following, we deal with a situation where the region of interest lies within the $n_i=1$ Mott-insulating lobe and the local zero-temperature state, in the atomic limit, is described by $|\Psi\rangle_{0,J/U=0}=\prod_i  |n_i=1\rangle$. The emergence of quantum-correlated particle-hole pairs can then be understood within first-order perturbation theory considering the tunneling term as a perturbation. To first order, one obtains for the ground state
\begin{equation}\label{eq:PertTheory}
  |\tilde \Psi \rangle_{0}
\propto
   |\Psi\rangle_{0,J/U=0} + \frac{J}{U}\sum_{\langle i,j \rangle} \hat a_i^\dagger \hat a_j |\Psi\rangle_{0,J/U=0}.
\end{equation}
The second term yields contributions of the schematic form $\hat a_i^\dagger \hat a_j |\Psi\rangle_{0,J/U=0}=\sqrt{2}|1,1,...,0,2,...,1,1\rangle$, with a sum over all positions and orientations of the particle-hole pair formed by the neighboring empty- and doubly-occupied sites. The state $ |\tilde \Psi \rangle_{0}$ yields a probability to find a particle-hole pair on neighboring sites proportional to $(J/U)^2$.\\
Closer to the transition to the superfluid phase, higher-order perturbation terms become more  important. Intuitively, this leads to a rapid increase of bound particle-hole pairs and an extension of their size, eventually resulting in deconfinement of the pairs at the transition point. One might view this proliferation and extension of particle-hole pairs as the driving force for the superfluid-Mott-insulator transition. However, more complicated clusters of particles might also play an important role.\\  
We would like to stress that $ |\tilde \Psi \rangle_{0}$ specifies the ground state of the system. Particle-hole pairs can therefore be regarded as virtual excitations. This should be distinguished from actual excited states of  the Mott-insulating phase, which are reached by increasing or decreasing the atom number by one.
\subsubsection{Two-site parity-correlation functions}
For the detection of correlations induced by particle-hole pairs, we exploited the fact that they have a finite spatial extension. As a consequence, the appearance of a number fluctuation on a given site leads to an enhanced probability to find a fluctuation on a close-by site. This behavior is captured in a  two-site parity-correlation function \cite{Kapit:2010, Endres:2011}
\begin{equation}
	C_p(d) = \avs{\hat s_k \hat s_{k+d}} - \avs{\hat s_k}\avs{\hat s_{k+d}}\label{eq:Cd}.
\end{equation}
In contrast to the general definition in Eq.\,\ref{equ:local_order}, we subtract the term $\avs{\hat s_k}\avs{\hat s_{k+d}}$, i.e., $C_p(d)$ is a connected correlation function or a cumulant. In this form, $C_p(d)$ (for $d\neq0$) vanishes for all many-body states $|\Psi\rangle=\prod_i |\psi_i\rangle$, that are simple products of on-site states $|\psi_i\rangle$, because such states lead to a factorization $\avs{\hat s_k \hat s_{k+d}} =\avs{\hat s_k}\avs{\hat s_{k+d}}$. Therefore, we expect no signal for the ground state of the atomic limit (Eq.\,\ref{atomic_limit_ground}) and for the ground state of the non-interacting limit ($U/J\approx0$), which is effectively described by a product of on-site coherent states \cite{Bloch:2008c}. Additionally, thermally induced number fluctuations do not cause a signal in the atomic limit because the density operator for the finite-temperature ensemble is a product of on-site operators \cite{Gerbier:2007b}.\\
In contrast, the perturbation theory result in Eq.\,\ref{eq:PertTheory} shows a many-body state that is a superposition of product states that have either no particle-hole pair or a particle-hole pair at different positions.  The nearest-neighbor parity-correlation function is then $C_p(d=1)=16 (J/U)^2+\mathcal{O}[(J/U)^4]$, which is directly proportional to the probability to find a particle-hole pair on neighboring sites.\\
However, $C_p(d)$ cannot be used to formulate an order parameter for the superfluid to Mott insulator transition because $C_p(d)$ shows a rapid decay with $d$ in both phases\,\cite{Endres:2011}, which means that $\avs{\hat s_k \hat s_{k+d}}\approx \avs{\hat s_k}\avs{\hat s_{k+d}}$ for large $d$. Therefore, both phases cannot be distinguished with $C_p(d)$, in contrast to the non-local order parameter that we will discuss in Sec.\,\ref{non_local}. We use $C_p(d=1)$ solely to detect quantum fluctuations that are introduced by particle-hole pairs in the Mott insulating phase.
\subsubsection{Experimental results} 
\begin{figure*}
\resizebox{1\textwidth}{!}{%
\includegraphics{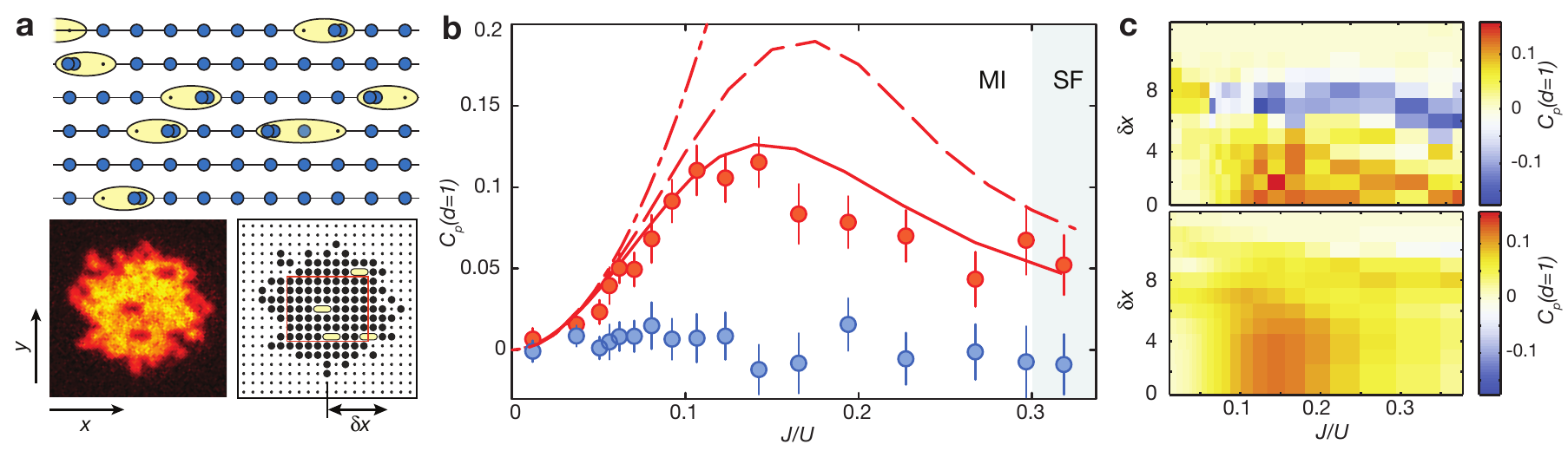}
}
\caption{\textbf{Nearest-neighbor correlations in 1d.} \textbf{a,} Top panel: Schematic image of the density distribution for $0<J/U\ll(J/U)_c$ with parallel 1d tubes and particle-hole pairs aligned in the direction of the tubes.  Bottom panel: Typical experimental fluorescence images for $J/U=0.06$ (left) and reconstructed on-site parity (right). Particle-hole pairs are emphasized by a yellow shading. \textbf{b,} 1d nearest-neighbor correlations $C_p(d=1)$ as a function of $J/U$ averaged over  a central region of $9 \times 7$ lattice sites (box in lower right panel of \textbf{a}).  The data along the tube direction ($x$-direction, red circles) shows a positive correlation signal, while the signal in orthogonal direction ($y$-direction, blue circles) vanishes within error-bars due to the decoupling of the 1d systems. The curves are first-order perturbation theory (dashed-dotted line), Density-Matrix Renormalization Group (DMRG) calculations for a homogeneous system at temperature $T=0$ (dashed line) and finite-temperature Matrix Product State (MPS) \cite{Verstraete:2004,Zwolak:2004} calculations including harmonic confinement at $T=0.09\,U/k_B$ (solid line). The error bars denote the 1$\sigma$ statistical uncertainty. The light blue shading highlights the superfluid phase. \textbf{c,} $C_p(d=1)$ as a function of the distance $\delta x$ from the center of the 1d tubes (as shown by the arrow in the lower right panel of \textbf{a}). The experimental data (top panel) shows a positive correlation signal for intermediate $J/U$ and for $\delta x \lesssim 4$ in accordance with the theoretical prediction based on MPS simulations (bottom panel). We attribute a negative experimental signal at the edge of the cloud ($\delta x \approx 8$) to shot-to-shot fluctuations of the cloud size. For the experimental data in \textbf{c}, we averaged the central $7$ tubes and additionally averaged over data points with the same distance $\delta x$ to the left and the right of the center of the 1d tubes. The signal in \textbf{b} stems from an average over $\delta x \leq 4$ of the signal in \textbf{c}.  Parts of Fig. \textbf{a}, and Fig.\,\textbf{b} are from Ref.\,\cite{Endres:2011}.}
\label{fig:3}       
\end{figure*}

We analyzed two-site parity correlations in 1d systems. For this, we split the 2d quantum gas into parallel 1d systems by choosing a high final lattice depth in $y$-direction of $V_y\approx 17 E_r$ effectively inhibiting tunneling in this direction (see sequence in Fig.\,\ref{fig:1}b and illustration in Fig.\,\ref{fig:3}a). We varied $J/U$ within these 1d systems by varying the final lattice depth in the $x$-direction from $V_x\approx 4\,E_r$ to $V_x \approx 17\,E_r$. Due to the quasi-adiabatic preparation, we expect the system to be close to the equilibrium state for the final $J/U$ values, \\
Typically, our samples  contained  $150 - 200$ atoms in order to avoid  Mott insulators of occupation numbers $n_i>1$. For small $J/U$ values, we therefore probed the correlations within the $n_i=1$ Mott lobe. We first evaluated $C_p(d=1)$ for each nearest-neighbor pair of sites in a central region of $9 \times 7$ lattice sites by an ensemble average over $50-100$ experimental repetitions. We then performed a spatial average over this central region. The spatially averaged signal can be written as $\sum_{k\epsilon R}(\avs{\hat s_k \hat s_{k+1}} - \avs{\hat s_k}\avs{\hat s_{k+1}})/N_R$, where $N_R$ is the number of nearest-neighbor pairs of sites in the central region $R$.\\
First, we recorded the nearest-neighbor correlations $C_p(d=1)$ for different values of $J/U$ along the direction of the 1d tubes (red circles in Fig.\,\ref{fig:3}b). For $J/U\approx 0$, the nearest-neighbor correlations vanish within errorbars, compatible with the product-state description of the many-body state in this limit.  As particle-hole pairs emerge with increasing $J/U$, we observe an increase of nearest-neighbor correlations. The appearance of a maximum in the signal can be understood from the fact that in the non-interacting limit ($U/J=0$), $C_p(d)$ also has to vanish.  The precise location and value of this maximum is, however, hard to predict. Note that the maximum is reached for a $J/U$-value that is higher than the critical value in mean-field theory $(J/U)_{c,{\rm MF}}^{1d} \approx 0.09$ but lower than the more precise critical value $(J/U)_c^{1d}\approx 0.3$ obtained with numerical methods \cite{Kashurnikov:1996b, kuehner:2000}. For a discussion of the comparison with the numerical calculation shown in Fig.\,\ref{fig:3}b, we refer to Ref.\,\cite{Endres:2011}. Furthermore, we show $C_p(d=1)$ as a function of the distance from the center of the 1d tubes (see Fig.\,\ref{fig:3}c and figure caption). \\
In addition to the nearest-neighbor parity correlation $C_p(d=1)$, we also evaluated 
$C_p(d)$ for distances $d=0$ and $d=2$ in our 1d systems (see supplementary material of Ref.\,\cite{Endres:2011}). In contrast to $C_p(d=1)$, the on-site parity fluctuations $C_p(d=0)=\avs{\hat s_k^2} - \avs{\hat s_k}^2$ do not completely vanish in the atomic limit due to thermally induced fluctuations, and $C_p(d=0)$ increases monotonically with increasing $J/U$ saturating at $C_p(d=0)\approx 1$ in the superfluid regime. As a function of the distance $d$, $C_p(d)$ drops rapidly and the numerical calculations predict only a small maximum of $0.01$ in the next-nearest-neighbor correlation $C_p(d=2)$ at $J/U\approx 0.17$, which is indiscernible from the statistical noise in our measurements \cite{Endres:2011}.\\
\subsection{Detection of non-local correlations in the Bose-Hubbard model}
\label{non_local}
\begin{figure}
\resizebox{0.5\textwidth}{!}{%
\includegraphics{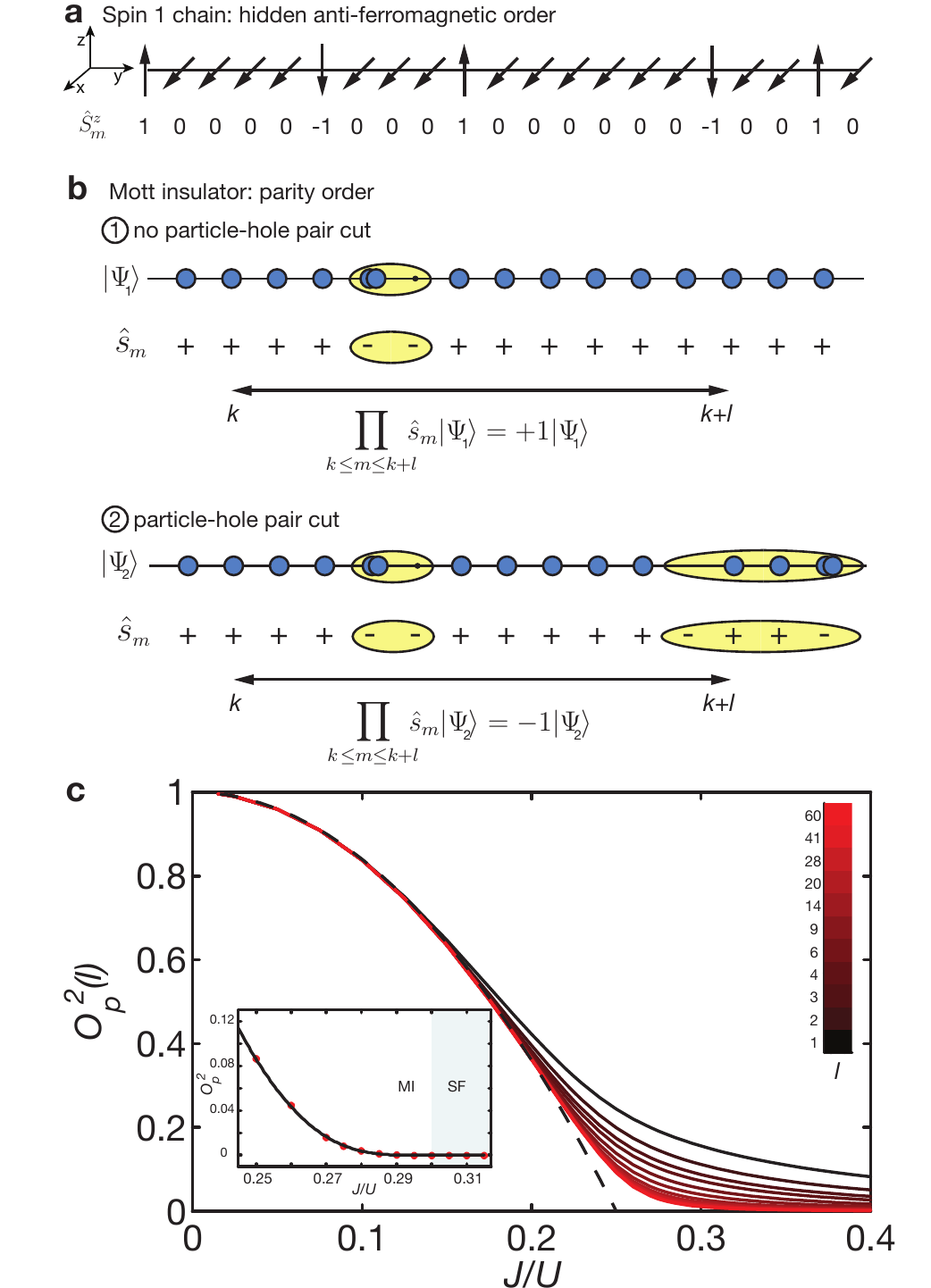}
}
\caption{\textbf{Illustration of non-local order and numerical results. a}, Hidden anti-ferromagnetic order in a spin-1 chain. Each spin with $z$-component $+1$ is followed by a spin with $-1$, with an arbitrary number of spins with $0$ in between. This order is hidden for a two-site correlation function, but can be detected using the non-local correlation function in Eq.\,\ref{string_spin1}. {\bf b}, Illustration of parity order in 1d Mott insulators. For states where all particle-hole pairs lie within the evaluation region $[k, k+l]$ , the string operator $\prod_{k\leq m \leq k+l} \hat{s}_m$ has an eigenvalue of $+1$ because the introduced pairwise minus signs cancel (upper panel). Only states which have a particle-hole pair extending out of the evaluation region yield $-1$ (lower panel). \textbf{c,} ${\cal O}_p^2(l)$ as a function of $J/U$ calculated with DMRG for a homogeneous chain  ($\bar n = 1$, $T = 0$) of total length $216$. Lines show ${\cal O}_p^2(l)$ for selected lengths $l$ (black to red colors). The dashed line shows the first-order strong coupling result ${\cal O}_p^2(l)=1-16(J/U)^2$.  Inset: Extrapolated value ${\cal O}_p^2=\lim_{l\rightarrow\infty}{\cal O}_p^2(l)$ together with a fit (black line) of the form  \mbox{$ {\cal O}_p^2\propto \exp \big(-A\left [(J/U)_c-(J/U)\right ]^{-1/2}\big)$}, which is characteristic for a transition of  Berezinskii-Kosterlitz-Thouless type. Fig. \textbf{c} without the strong coupling result was first published in Ref.\,\cite{Endres:2011}.}
\label{fig:4}       
\end{figure}
\subsubsection{Perturbation analysis} 
For the following discussion of non-local order in the Mott insulating phase, let us define a string operator
\begin{align}
\hat{O}_p(l)=\prod_{k\leq m \leq k+l}\hat s_m,
\end{align}
and the following notation  for the string correlator
\begin{align}
{\cal O}_p^2(l)=\langle \hat{O}_p(l)\rangle.
\end{align}
With this, there is string order if
\begin{align}
{\cal O}_p^2=\lim_{l \rightarrow \infty}  {\cal O}_p^2(l) >0. 
\end{align}
In the atomic limit, at $T=0$, the Mott-insulating ground state has the form $|\Psi\rangle_{0,J/U=0}=\prod_i |n_i=1\rangle$. For this state, one finds trivially $\langle \hat{O}_p(l) \rangle=1$ and therefore string order. We will now demonstrate that string order is stable with respect to finite tunneling $J/U>0$, based on the perturbation expansion in Eq.\,\ref{eq:PertTheory}. All states that are summed up in the expansion are eigenstates of the local parity operator $\hat{s}_m$. If no defect is encountered at site $m$, $\hat{s}_m$ has an eigenvalue $s_m=+1$ and, if a defect is encountered, we find $s_m=-1$. Therefore, the states in the expansion on the right hand side of Eq.\,\ref{eq:PertTheory} are eigenstates of $\hat{O}_p(l)$ with eigenvalues $\prod_{k\leq m \leq k+l} s_m=\pm 1$.\\
The sign of the eigenvalue depends on whether or not a particle-hole pair is cut at the ends of the string operator $\hat{O}_p(l)$. By this, we mean the following: Consider a state $\hat a_i^\dagger \hat a_j |\Psi\rangle_{0,J/U=0}$ which has a particle-hole pair at positions $i$ and $j$, where $i$ is the position of the particle and $j$ the position of the hole. A particle-hole pair is cut if either $i$ or $j$ is within the interval $[k,k+l]$ that is sampled by the string operator. If a single pair is cut, the eigenvalue of $\hat O_p(l)$ is $-1$, because we find an unpaired minus sign. In all other cases, the eigenvalue is $+1$ because particle-hole pairs inside $[k,k+l]$ lead to pairwise minus signs that cancel (Fig.\,\ref{fig:4}b).\\
In the expansion of Eq.\,\ref{eq:PertTheory}, we find four different contributions for which such a cut is possible because there are two endpoints of the string operator and two permutations of particle and hole. Each of these contributions appears with a quantum mechanical amplitude of $\sqrt{2}J/U$. The total probability to cut a particle-hole pair is therefore $p_c=4 (\sqrt{2} J/U)^2=8 (J/U)^2$ and the total probability not to cut a particle-hole pair is $p_{nc}=1-8 (J/U)^2$. Based on the argument of the previous paragraph, we  can write
\begin{align}
{\cal O}_p^2(l)=p_{nc}-p_c=1-16 (J/U)^2.
\end{align}
This result is valid if $l\geq 2$ and for $J/U\ll (J/U)_c$. Consequently, we find ${\lim_{l \rightarrow \infty} \langle {\cal O}_p^2(l) \rangle=1-16 (J/U)^2>0}$. Therefore, Mott insulators in 1d at small $J/U$ possess string order.\\
The previous analysis is expected to hold in a similar way in the whole Mott phase if higher order terms in the perturbation series are included. The crucial point is that for a given order $(J/U)^n$ in the Mott-insulating phase, defects appear in finite-sized clusters with a typical extension $l_c$. For low orders of $n$, we expect $l_c\approx n$. Within such clusters, the number of lattice sites with eigenvalue $s_m=-1$ is always even because the application of the perturbation operator $\hat a_i^\dagger a_j$, with $i,j$ being neighboring sites, can only change this number by $\pm2$. For example, the application of $\hat a_i^\dagger a_j$ on a state $|1,1,...,1,0,2,1..,1,1\rangle$ can create another particle-hole pair, destroy the particle-hole pair, or lead to states $|1,1,...,0,1,2,...,1,1\rangle$ or $|1,1,...,0,3,0,...,1,1\rangle$. In all cases, we find an even number of sites with $s_m=-1$. If such a finite-sized cluster lies completely within the evaluation region $[k,k+l]$, it contributes only with an overall plus sign since the contribution of all sites with $s_m=-1$ cancels pairwise. The string-correlation function ${\cal O}_p^2(l)$ then takes on a constant value if $l \gg l_c$ because clusters can only be cut at the ends of the evaluation region and the probability for such a cut stays constant as a function of the length $l$.
\subsubsection{Numerical analysis}
This intuitive view is supported by a numerical analysis using DMRG (Fig.\,\ref{fig:4}c, solid lines). The result from the strong coupling expansion ${\cal O}_p^2\approx 1-16 (J/U)^2$ (dashed line) agrees with the DMRG data for small $J/U$ values. The inset of Fig.\,\ref{fig:4}c shows the extrapolated values ${\cal O}_p^2=\lim_{l\rightarrow\infty}{\cal O}_p^2(l)$, which we calculated using finite size scaling (see supplementary material of Ref.\,\cite{Endres:2011} for details). We observe that ${\cal O}_p^2>0$ in the Mott phase and that ${\cal O}_p^2$ vanishes at $J/U\approx 0.3$, which is compatible with the known numerical value of $(J/U)_c\approx 0.3$ for the critical coupling strength for the 1d superfluid-Mott-insulator transition \cite{Kashurnikov:1996b, kuehner:2000}. This analysis shows that 1d bosonic Mott insulators possess string order for all coupling strengths $J/U<(J/U)_c$.\\
Without going into details, we note that an analysis using a Bosonization treatment yields an algebraic decay to zero of ${\cal O}_p^2(l)$ with $l$, in the superfluid phase, and an approximately exponential decay to a constant non-zero value, in the Mott-insulating phase \cite{Mazza:2012}. Furthermore, a mapping to a 2d interface model shows that ${\cal O}_p^2$ grows as {$ {\cal O}_p^2\propto \exp \big(-A\left [(J/U)_c^{1d}-(J/U)\right ]^{-1/2}\big)$} at the transition point \cite{Rath:2013}, and our numerical data is compatible with this behavior (inset of Fig.\,\ref{fig:4}c). The 1d superfluid-Mott-insulator transition is of  Berezinskii-Kosterlitz-Thouless type, and the latter scaling is the same scaling as for the opening of the Mott-insulating gap at the transition point \cite{Kuehner:1998}.
\begin{figure}
\resizebox{0.5\textwidth}{!}{%
\includegraphics{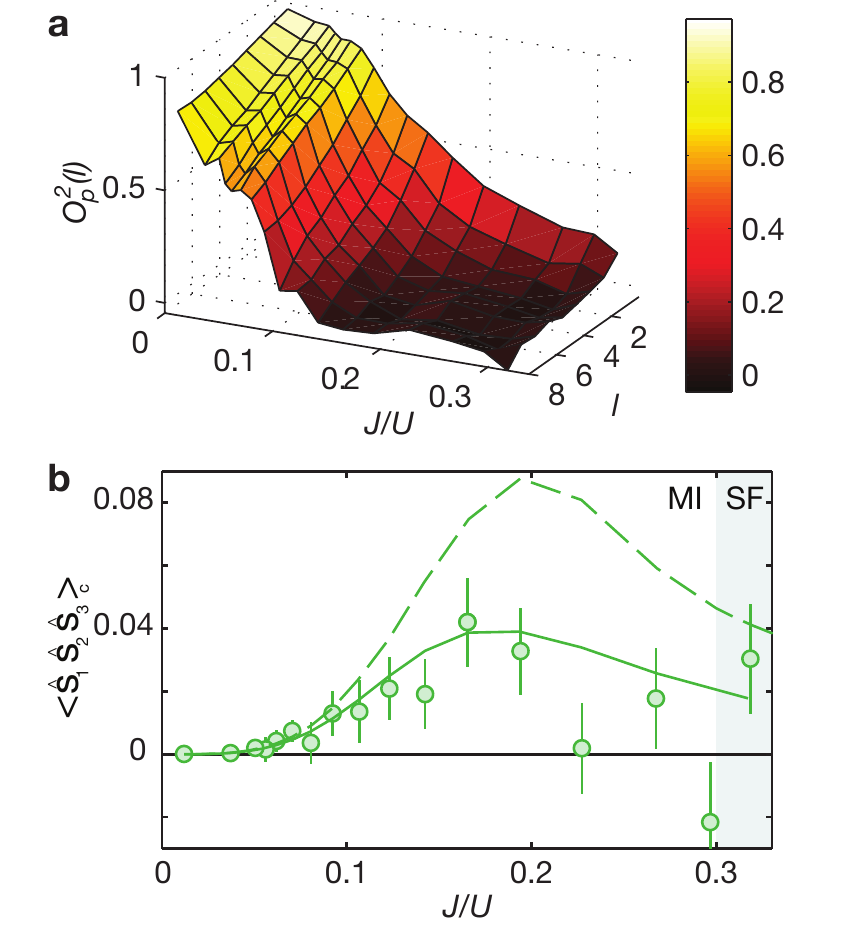}
}
\caption{\textbf{Experimental results for non-local correlations. a}, Experimental values of ${\cal O}_p^2(l)$ for lengths $0 \leq l \leq 8$  \textbf{b,} Our experimental values for three site cumulant $\avs{\hat{s}_1\hat{s}_2\hat{s}_3}_c$ (green circles) show the existence of three-site correlations in our system.  The curves are DMRG calculations for a homogeneous system at $T=0$ (dashed line) and finite-temperature MPS calculations including harmonic confinement at $T=0.09\,U/k_B$ (solid line). Figs. \textbf{a,b} are from Ref.\,\cite{Endres:2011}.}
\label{fig:5}       
\end{figure}
\subsubsection{Experimental results} Our experimentally obtained values of  ${\cal O}_p^2(l)$ for string lengths $l \leq 8$ are shown in Fig.\,\ref{fig:5}a. We observe a stronger decay of ${\cal O}_p^2(l)$ with $l$ compared to the $T=0$ case because at finite temperature, thermal fluctuations lead to minus signs at random positions and reduce the average value of ${\cal O}_p^2(l)$. Despite this, we still see a strong growth of ${\cal O}_p^2(l)$ once the transition from the superfluid to the Mott insulator is crossed, similar to the behavior in Fig.\,\ref{fig:4}c. For a more detailed comparison with numerical results, we refer to Ref.\,\cite{Endres:2011}.\\
Here we would like to stress that this signal cannot be explained with pure two-site correlations. To illustrate this, let us consider the simplest case of a string-type correlator including three sites $\avs{\hat{s}_1\hat{s}_2\hat{s}_3}$, where $\hat{s}_1$,$\hat{s}_2$ and $\hat{s}_3$ refer to the parity of three neighboring sites on a 1d chain. Three-site correlations are captured by a three-site cumulant $\avs{\hat{s}_1\hat{s}_2\hat{s}_3}_c$, defined as \cite{Kubo:1962}
\begin{align}
\avs{\hat{s}_1\hat{s}_2\hat{s}_3}_c= &\avs{\hat{s}_1\hat{s}_2\hat{s}_3}-\avs {\hat{s}_1}\avs{\hat{s}_2}\avs{\hat{s}_3}\nonumber \\&-C_p(1,2)\avs{\hat{s}_3}-C_p(2,3)\avs{\hat{s}_1} -C_p(1,3)\avs{\hat{s}_2} \label{eq:fullthree},
\end{align}
with two-site correlation functions $C_p(i,j)=\avs{\hat{s}_i\hat{s}_j}-\avs{\hat{s}_i}\avs{\hat{s}_j}$. The three-site cumulant $\avs{\hat{s}_1\hat{s}_2\hat{s}_3}_c$ is a measure of the correlations between all three sites as it vanishes if one of the sites is not correlated with the others \cite{Kubo:1962}.\\ If $\avs{\hat{s}_1\hat{s}_2\hat{s}_3}_c$ vanishes,  Eq.\,\ref{eq:fullthree} can be written as $\avs{\hat{s}_1\hat{s}_2\hat{s}_3}=\avs {\hat{s}_1}\avs{\hat{s}_2}\avs{\hat{s}_3}+C_{1,2}\avs{\hat{s}_3}+C_{2,3}\avs{\hat{s}_1} +C_{1,3}\avs{\hat{s}_2}$. In this case, $\avs{\hat{s}_1\hat{s}_2\hat{s}_3}$ does not contain more information than two-site and on-site terms. In contrast, a non-zero $\avs{\hat{s}_1\hat{s}_2\hat{s}_3}_c$ signals that $\avs{\hat{s}_1\hat{s}_2\hat{s}_3}$ contains additional information. Our experimental values for the three-site cumulant $\avs{\hat{s}_1\hat{s}_2\hat{s}_3}_c$ (Fig.\ref{fig:5}b) show a significant signal for $J/U<0.20$, in quantitative agreement with the MPS calculation. \\
In summary, for $J/U \approx 0$, the string-correlation signal is purely dominated by on-site terms while for ${0 \lesssim J/U \lesssim  0.1}$, also two-site correlations (quantified by the cumulant in Eq.\,\ref{eq:Cd}) start to play a role. In the range $0.1 \lesssim J/U \lesssim  0.2$, three-site correlations (quantified by the cumulant in Eq.\,\ref{eq:fullthree}) build up and also contribute to the string-correlation signal. In general, one expects higher-order correlations to play a more significant role in the vicinity of the phase transition. A study of cumulants including more than three sites is a matter of future investigations.
\section{Discussion and outlook}
\label{conclusion}
We presented experimental and theoretical details for the single-atom and single-site-resolved detection of correlation functions in a strongly interacting 1d lattice system \cite{Endres:2011}. More specifically, we studied two-site and non-local parity correlations across the 1d superfluid-Mott-insulator transition.\\
In Ref.\,\cite{Endres:2011}, we also showed the detection of two-site correlation functions across the 2d transition. It is an interesting question whether non-local correlation functions can also be used to detect the phase transition in 2d using a function ${\cal O}_p^2(A)=\langle \prod_{i\epsilon A} \hat s_i \rangle$ that evaluates the product of on-site parities on an area $A$. In contrast to the 1d situation, the boundary $\partial A$ of the evaluation area grows when increasing the area. One can show that even for first order in $J/U$, this leads to an exponential decay $\log[{\cal O}_p^2(A)]\propto-\partial A$ in the Mott-insulating phase. However, the superfluid phase is expected to show a faster decay with $\log[{\cal O}_p^2(A)]\propto-\partial A \log(\partial A)$, and the phase transition should be detectable by these different scalings \cite{Rath:2013}.\\
Furthermore, the detection of single-atom and single-site-resolved correlation functions is a valuable tool for the study of out-of-equilibrium behavior. In Ref.\,\cite{Cheneau:2012}, we could detect the spreading of correlations after a quench in the 1d Mott insulating regime. It would be interesting to extend these studies to 2d systems, where the numerical calculation of out-of-equilibrium dynamics is demanding.\\
Another future direction is the single-site and single-atom-resolved detection of ultracold fermions in the Fermi-Hubbard regime. The measurement of correlations could then be used for a direct detection of anti-ferromagnetic order at low temperatures.\\
Additionally, in a recent experiment, we used the same technique to detect correlations between Rydberg atoms forming spatially ordered structures \cite{Schauss:2012}, showing that the general approach is not restricted to Hubbard-type physics. Furthermore, using off-resonant Rydberg-dressing \cite{Honer:2010, Pupillo:2010}, an extended Bose-Hubbard model with longer-range interactions that features a Haldane insulating phase \cite{DallaTorre:2006,Berg:2008} could be realized. This phase is expected to show a non-local order similar to the hidden anti-ferromagnetic order in Eq.\,\ref{string_spin1}.\\
The possibility to measure higher-order correlation functions offers a precise way of characterizing different quantum phases. In the long term, however, one wishes to perform measurements that are not restricted to density-type correlation functions in order to detect coherences in the many-body state (see discussion in Sec.\,\ref{observable}). Therefore,  a major experimental step would be the single-site-resolved detection of the single-particle density matrix $\langle  \hat{a}^\dagger_i \hat{a}_j\rangle$. Based on this, a long-term goal would be the direct measurement of correlators involving higher orders of creation and annihilation operators that do not resemble density correlations. One particularly interesting application is the investigation of entanglement in many-body systems \cite{Amico:2008}. Indeed, recent proposals suggested that single-site- and single-atom-resolved imaging could be used for the detection of entanglement in bosonic \cite{Daley:2012} and fermionic \cite{Pichler:2013} systems. Possible applications range from the detection of entanglement growth after quantum quenches to experimental investigations of area laws.
\section*{Acknowledgments} We thank Jacob Sherson for his contribution to the experimental setup. We acknowledge helpful discussions with Ehud Altman, Emanuele Dalla Torre, Matteo Rizzi, Ignacio Cirac, Andrew Daley, Peter Zoller, Steffen Patrick Rath, Wolfgang Simeth and Wilhelm Zwerger. This work was supported by MPG, DFG, EU (NAMEQUAM, AQUTE, Marie Curie Fellowship to M.C.), and JSPS (Postdoctoral Fellowship for Research Abroad to T.F.). LM acknowledges the economical support from Regione Toscana, POR FSE 2007-2013. DMRG simulations were performed using code released within the PwP project (\url{www.qti.sns.it}).

\bibliographystyle{bib/manubib2}
\bibliography{bib/References}
\end{document}